# Theory of the Chromatic Dispersion, Revisited

Dimitar Popmintchev[1]✉, Siyang Wang[2], Xiaoshi Zhang[2], Tenio Popmintchev[1,2]✉

[1]*Photonics Institute, Vienna University of Technology, Vienna A-1040, Austria*

[2]*University of California at San Diego, CO 92309-0440 USA*

We derive general analytic expressions for the chromatic dispersion orders valid to infinity, due to the *k* vector or phase *φ* dependence on the wavelength. Additionally, we identify polynomials and recursion relations associated with the chromatic dispersion orders and draw analogy to the generalized Lah and Laguerre transformations. Further, we give explicitly the dispersion terms to the 10$^{th}$ order and visualize the chromatic dispersion for material, grating and prism- pair compressors and hollow - core photonic anti-resonant fiber. These simple formulas are applicable for material dispersion, compressors, stretchers, waveguides, and any other type of known frequency-dependent phase.

Often assumed negligible, the high chromatic dispersion orders are commonly avoided from discussions. Here we delve once more into this longstanding phenomenon, dating back to the studies of Pierre Laplace and George Airy on the dispersion of the water waves, and Ernst Abbe on optical systems, where we are interested in obtaining a general closed functional form of the chromatic dispersion orders and further unveiling their significance in view of the effect they have on ultrashort pulses. To the best of our knowledge, such generalization has not been attained previously.

The chromatic dispersion phenomenon has worked both in-favor and against some of the greatest novelties of the past and present century. The group velocity dispersion (GDD) and the higher orders of dispersion have been a major hindrance for the fiber-based telecommunication. With the ever-growing need of increasing the transferring bit-rate length product through these hair-thin fibers, the technology has been pushed to its limits. On one hand, pulsed signals were desired to transfer the information and on the other, these pulses were needed to preserve their structure to an extent that the receiver can decode the message. Using a long narrow-band pulses at the zero GDD dispersion wavelength was a compromise [1, 2, 3, 4, 5], considering that even a lengthy sub-$ps$, narrow-band pulse could spread beyond recognition, when propagating few tens of kilometers. Nevertheless, the dispersion can be manipulated by sections of dispersion- shifted fibers, Bragg gratings, etc. However, operating away from the zero dispersion is advantageous for increased data transmission through multiple channels

✉e-mail: *dimitar.popmintchev@gmail.com*, *tenio.popmintchev@physics.ucsd.edu*

(wavelength division multiplexing), where some residual dispersion is needed to avoid signal distortions [6].

In another area, similarly to the radar detection systems, the chirp pulse amplification technique (CPA), where the pulses are at first stretched in the time domain to prevent damage of the materials, then amplified and lastly compressed to ultrashort levels, has opened the gateways to the development of immensely high power and energy laser systems at high pulse repetition rate [7, 8, 9]. At the same time achieving a near transform-limited ultrashort pulse duration required a nearly impossible balance of the chromatic dispersion acquired from the stretcher and the consequent laser amplification stages where the pulses accumulate linear and nonlinear phases passing through lengthy materials. While 'matching' - grating and/or - prism compressors and stretchers [10, 11] are suitable for compensation for all chromatic orders, it is not always possible to adequately cancel the higher orders due to linear or non-linear propagation through the materials. Then specially designed multilayered thin-film structures could be of use to further compensate the pulses' dispersions to some extent at the cost of reduced reflectivity or transmission [12, 13, 14].

In order to craft a more comprehensive picture of the chromatic dispersion, we start from the basics. Any well-behaved wave function can be represented by sinusoidal wavelets having different phase velocity, in a dispersive medium, that depends on the frequency. The propagation of such a wave in dispersive systems is a linear and causal phenomenon that leads to pulse distortions. In linear systems the dispersion relation for the phase can be written as:

$$\varphi(\omega|\lambda) = k(\omega)z = \frac{\omega}{c}n(\omega)z = \frac{2\pi}{\lambda}n(\lambda)z = \frac{\omega}{c}OP(\omega) = \frac{2\pi}{\lambda}OP(\lambda) = \omega\tau(\omega) = \frac{2\pi}{\tau_0}\tau(\omega)$$

Where $OP(\lambda)$ is the optical path, $n(\omega)$ is the refractive index for the medium under consideration, $\tau(\omega)$ is the corresponding temporal interval and $\tau_0$ is the single- cycle pulse duration for wavelet with wavelength $\lambda$. We consider intensities where the interactions are linear. In such picture the signal spectral content will not change, but will be rearranged temporally over a certain band around an average frequency $\omega_0$. As a convenience, the spectral phase can be expanded around this average frequency in Taylor series [15]:

$$\varphi(\omega) = \varphi|_{\omega_0} + \frac{\partial \varphi}{\partial \omega}\bigg|_{\omega_0}(\omega - \omega_0) + \frac{1}{2}\frac{\partial^2 \varphi}{\partial \omega^2}\bigg|_{\omega_0}(\omega - \omega_0)^2 + \cdots + \frac{1}{p!}\frac{\partial^p \varphi}{\partial \omega^p}\bigg|_{\omega_0}(\omega - \omega_0)^p + \cdots$$

$$= \varphi|_{\omega_0} + \tau_g|_{\omega_0}(\omega - \omega_0) + \frac{1}{2}GDD(\omega - \omega_0)^2 + \cdots + \frac{1}{p!}POD(\omega - \omega_0)^p + \cdots$$

The first derivative $\frac{\partial \varphi}{\partial \omega}\bigg|_{\omega_0} = \frac{\partial \tau}{\partial \omega} = \tau_g$ is the group delay (GD). It results in a shift of the envelope of the pulse in the temporal space. The higher orders cause changes of the temporal pulse shape. The second term $\frac{\partial^2 \varphi}{\partial \omega^2}\bigg|_{\omega_0} = \frac{\partial}{\partial \omega}\tau_g(\omega)\bigg|_{\omega_0}$ is the group delay dispersion (GDD) and

in general $\left.\frac{\partial^p \varphi}{\partial \omega^p}\right|_{\omega_0} = \left.\frac{\partial^{p-1}}{\partial \omega^{p-1}} \tau_g(\omega)\right|_{\omega_0}$ is the $p$ order dispersion (POD). This description requires knowledge of the particular $\tau_g(\omega)$ and can give recursive expressions for special functions with repetitive derivatives [16]. Inverse chain rules are needed to estimate the chromatic dispersion when $\omega(\tau_g)$ is easier to be specified, as it is in the case for the chromatic dispersion of the high harmonic generation process [17, 18].

In this paper, we are interested in closed form general expressions when the optical path $OP(\lambda)$ or the refractive index $n(\lambda)$ is known. In the same manner, the wave vector $k(\omega|\lambda)$ or optical path $OP(\omega|\lambda)$ can be expanded in Taylor series [15]:

$$k(\omega) = k|_{\omega_0} + \left.\frac{\partial k}{\partial \omega}\right|_{\omega_0}(\omega - \omega_0) + \frac{1}{2}\left.\frac{\partial^2 k}{\partial \omega^2}\right|_{\omega_0}(\omega - \omega_0)^2 + \cdots + \frac{1}{p!}\left.\frac{\partial^p k}{\partial \omega^p}\right|_{\omega_0}(\omega - \omega_0)^p =$$

$$= k_0 + v_{gr}^{-1}(\omega - \omega_0) + \frac{1}{2}GDD(\omega - \omega_0)^2 + \cdots + \frac{1}{p!}POD(\omega - \omega_0)^p + \cdots$$

The lowest term $\left.\frac{\partial k}{\partial \omega}\right|_{\omega_0}$ is the inverse group velocity. The second term $\left.\frac{\partial^2 k}{\partial \omega^2}\right|_{\omega_0}$ is the group delay dispersion (GDD) and in general $\left.\frac{\partial^p k}{\partial \omega^p}\right|_{\omega_0}$ is the $p$ order dispersion (POD). When the pulses have substantial spectral bandwidth, it is necessary to take into account also the higher orders in the Taylor expansion.

The calculation of the chromatic dispersion terms is straight forward in the frequency domain, where we take the successive derivatives of the wave vector $k(\omega)$ or the phase $\varphi(\omega)$. The first few terms are given explicitly in Appendix A and C. In general:

$$\frac{\partial^p}{\partial \omega^p} k(\omega) = \frac{1}{c}\left(p \frac{\partial^{p-1}}{\partial \omega^{p-1}} n(\omega) + \omega \frac{\partial^p}{\partial \omega^p} n(\omega)\right) \quad (1)$$

$$\frac{\partial^p}{\partial \omega^p} \varphi(\omega) = \frac{1}{c}\left(p \frac{\partial^{p-1}}{\partial \omega^{p-1}} OP(\omega) + \omega \frac{\partial^p}{\partial \omega^p} OP(\omega)\right) \quad (2)$$

Where $p = 1, 2, 3, 4 \ldots$ are integer numbers.

In terms of wavelength, the above expressions are much more complex. Starting from the derivative of the angular frequency $d\omega = -\frac{2\pi c}{\lambda^2} d\lambda$, we can evaluate consecutively each of the derivatives of $n(\omega)$ or $OP(\omega)$ in terms of the wavelengths. Despite of the complexity, the frequency derivatives simplify to polynomials of order $p$ in $\lambda$:

$$\frac{\partial^p}{\partial \omega^p} n(\omega) = (-1)^p \left(\frac{\lambda}{2\pi c}\right)^p \sum_{m=0}^{p} \mathcal{A}(p,m) \lambda^m \frac{\partial^m}{\partial \lambda^m} n(\lambda) \quad (3)$$

Where $p = 1,2,3,4 \ldots$ are integer numbers and the matrix $\mathcal{A}(p,m)$ is given by:

$$\mathcal{A}(p,m) = \frac{p!}{(p-m)!m!} \frac{(p-1)!}{(m-1)!} = C(p,m)\frac{(p-1)!}{(m-1)!} = C(p-1, p-m)\frac{p!}{m!}$$

Where $C(p,m) = \binom{p}{m} = \frac{p!}{(p-m)!m!}$ are the binomial coefficients. The matrix elements $\mathcal{A}(p,m)$ are given in Table 1 up to the 10$^{th}$ order $p \leq 10$. At $m = 0$ we have $\mathcal{A}(p,0) = 0$ for $p \geq 1$. The derivatives are given explicitly up to the 10$^{th}$ order in Appendix B.

Similarly, we can write:

$$\frac{\partial^p}{\partial \omega^p} OP(\omega) = (-1)^p \left(\frac{\lambda}{2\pi c}\right)^p \sum_{m=0}^{p} \mathcal{A}(p,m) \lambda^m \frac{\partial^m}{\partial \lambda^m} OP(\lambda) \qquad (4)$$

Further, we note, that the matrix elements $\mathcal{A}(p,m)$ denote, amongst other known subseries, the values of the Lah numbers [19, 20]. In this sense, the derivative transformation to the wavelength space can be seen as a representation of the Lah transformation. The forward $\mathcal{L}(x)$ and inverse $\mathcal{L}^{-1}(u)$ Lah transform can be written as [21, 22]:

$$u_p = \mathcal{L}(x) = \sum_{m=0}^{p} \mathcal{A}(p,m) x_m; \quad x_p = \mathcal{L}^{-1}(u) = \sum_{m=0}^{p} (-1)^{p-m} \mathcal{A}(p,m) u_m$$

The generating function for the generalized polynomial $G_p^{(\alpha)}(x)$ containing the sums in equations (3) and (4) can be expressed for $p \geq 1$ and $\alpha = -1$, as:

$$G_p^{(\alpha)}(x) = x^{-\alpha} \frac{d^p}{dx^p}\left(x^{p+\alpha} f(x)\right) = \sum_{m=0}^{p} C(p+\alpha, p-m) \frac{p!}{m!} x^m f^{(m)}(x)$$

Where $f(x)$ is smooth $p$- times differentiable function representing the refractive index $n$ or the optical path $OP$, and $f^{(m)}(x)$ is the $m^{th}$ derivative of $f(x)$. Some of the properties of these polynomials are discussed in the next section.

Table 1. Matrix elements $\mathcal{A}(p,m)$ and Lah numbers up to the 10$^{th}$ order.

| Derivative Order $p$ | $d_\lambda^{(1)}$ $m=1$ | $d_\lambda^{(2)}$ $m=2$ | $d_\lambda^{(3)}$ $m=3$ | $d_\lambda^{(4)}$ $m=4$ | $d_\lambda^{(5)}$ $m=5$ | $d_\lambda^{(6)}$ $m=6$ | $d_\lambda^{(7)}$ $m=7$ | $d_\lambda^{(8)}$ $m=8$ | $d_\lambda^{(9)}$ $m=9$ | $d_\lambda^{(10)}$ $m=10$ |
|---|---|---|---|---|---|---|---|---|---|---|
| First $p=1$ | 1 | 0 | 0 | 0 | 0 | 0 | 0 | 0 | 0 | 0 |
| Second $p=2$ | 2 | 1 | 0 | 0 | 0 | 0 | 0 | 0 | 0 | 0 |
| Third $p=3$ | 6 | 6 | 1 | 0 | 0 | 0 | 0 | 0 | 0 | 0 |
| Fourth $p=4$ | 24 | 36 | 12 | 1 | 0 | 0 | 0 | 0 | 0 | 0 |
| Fifth $p=5$ | 120 | 240 | 120 | 20 | 1 | 0 | 0 | 0 | 0 | 0 |
| Sixth $p=6$ | 720 | 1800 | 1200 | 300 | 30 | 1 | 0 | 0 | 0 | 0 |
| Seventh $p=7$ | 5040 | 15120 | 12600 | 4200 | 630 | 42 | 1 | 0 | 0 | 0 |
| Eight $p=8$ | 40320 | 141120 | 141120 | 58800 | 11760 | 1176 | 56 | 1 | 0 | 0 |
| Ninth $p=9$ | 362880 | 1451520 | 1693440 | 846720 | 211680 | 28224 | 2016 | 72 | 1 | 0 |
| Tenth $p=10$ | 3628800 | 16329600 | 21772800 | 12700800 | 3810240 | 635040 | 60480 | 3240 | 90 | 1 |

Substituting the frequency derivatives (3) and (4) inside the frequency dispersion relations (1) and (2) we obtain much simpler expressions. Despite of the complexity, the dispersion orders simplify to polynomials of order $p$ in $\lambda$. The $p$ order chromatic dispersion is:

$$POD(n) = \frac{\partial^p}{\partial \omega^p} k(\omega) = (-1)^p \frac{1}{c}\left(\frac{\lambda}{2\pi c}\right)^{p-1} \sum_{m=0}^{p} \mathcal{B}(p,m) \lambda^m \frac{\partial^m}{\partial \lambda^m} n(\lambda) \qquad (5)$$

Where $p = 2, 3, 4, ...$ are integer numbers and the matrix $\mathcal{B}(p,m)$ is given by:

$$\mathcal{B}(p,m) = \frac{p!}{(p-m)!m!} \frac{(p-2)!}{(m-2)!} = \mathcal{A}(p,m)\frac{m-1}{p-1} = C(p,m)\frac{(p-2)!}{(m-2)!} = C(p-2, p-m)\frac{p!}{m!}$$

The matrix elements $\mathcal{B}(p,m)$ are given in Table 2 up to the $10^{th}$ order $p \leq 10$. At $m = 0, 1$ we have $\mathcal{B}(p, [0,1]) = 0$ for $p \geq 2$. The chromatic dispersion expressions are given explicitly up to the $10^{th}$ order in Appendix C.

Similarly, we can write:

$$POD(OP) = \frac{\partial^p}{\partial \omega^p} \varphi(\omega) = (-1)^p \frac{1}{c}\left(\frac{\lambda}{2\pi c}\right)^{p-1} \sum_{m=0}^{p} \mathcal{B}(p,m) \lambda^m \frac{\partial^m}{\partial \lambda^m} OP(\lambda) \qquad (6)$$

The refractive index $n(\lambda)$ and the optical path $OP(\lambda)$ are interchangeable in the chromatic dispersion equations, however, the units change. The $p^{th}$ order dispersion $POD(n)$ is measured in seconds to the power $p$ per unit length $[s^p/m]$, while the units of the total dispersion $POD(OP)$ are $[s^p]$. In fiber optics a scaled group-delay dispersion $SD = \frac{\partial \omega}{\partial \lambda} \frac{\partial \tau_g(\omega)}{\partial \omega}\bigg|_{\omega_0} = -\frac{\omega_0^2}{2\pi c} GDD$ is commonly used with units of $[s^2/m/nm]$, which represents the temporal separation between two wavelength-components after certain propagation length.

Table 2. Matrix elements $\mathcal{B}(p,m)$ up to the $10^{th}$ order and unsigned Laguerre coefficients for $\alpha = 2$.

| Dispersion order $p$ | $d_\lambda^{(2)}$ $m=2$ | $d_\lambda^{(3)}$ $m=3$ | $d_\lambda^{(4)}$ $m=4$ | $d_\lambda^{(5)}$ $m=5$ | $d_\lambda^{(6)}$ $m=6$ | $d_\lambda^{(7)}$ $m=7$ | $d_\lambda^{(8)}$ $m=8$ | $d_\lambda^{(9)}$ $m=9$ | $d_\lambda^{(10)}$ $m=10$ |
|---|---|---|---|---|---|---|---|---|---|
| Second $p=2$ | 1 | 0 | 0 | 0 | 0 | 0 | 0 | 0 | 0 |
| Third $p=3$ | 3 | 1 | 0 | 0 | 0 | 0 | 0 | 0 | 0 |
| Fourth $p=4$ | 12 | 8 | 1 | 0 | 0 | 0 | 0 | 0 | 0 |
| Fifth $p=5$ | 60 | 60 | 15 | 1 | 0 | 0 | 0 | 0 | 0 |
| Sixth $p=6$ | 360 | 480 | 180 | 24 | 1 | 0 | 0 | 0 | 0 |
| Seventh $p=7$ | 2520 | 4200 | 2100 | 420 | 35 | 1 | 0 | 0 | 0 |
| Eight $p=8$ | 20160 | 40320 | 25200 | 6720 | 840 | 48 | 1 | 0 | 0 |
| Ninth $p=9$ | 181440 | 423360 | 317520 | 105840 | 17640 | 1512 | 63 | 1 | 0 |
| Tenth $p=10$ | 1814400 | 4838400 | 4233600 | 1693440 | 352800 | 40320 | 2520 | 80 | 1 |

Further, we note that the matrix elements $\mathcal{B}(p,m) = \mathcal{A}(p,m)\frac{m-1}{p-1}$ ($m$ starting from 2), represent, amongst other known sub-series, the unsigned coefficients constructed as a triangular or square array $i!\left[C(i+2, i-j)\frac{1}{j!}\right]$ for the generalized orthogonal Laguerre polynomials $L_i^{(\alpha)}(x)$ for $\alpha = 2$ and shifted indices $i = p - 2, j = m - 2$ starting from 0.

The generating function for our generalized polynomial containing the sum in equations (5) and (6) can also be written in terms of the already defined polynomial $G_p^{(\alpha)}(x)$ for $p \geq 2$ and $m \geq 0$: $G_p^{(\alpha)}(x) = x^{-\alpha}\frac{d^p}{dx^p}(x^{p+\alpha}f(x)) = \sum_{m=0}^{p} C(p+\alpha, p-m)\frac{p!}{m!}x^m f^{(m)}(x)$ for $\alpha = -2$. For the Laguerre polynomials $f(x) = e^{-x}$, however, here $f(x) = e^{\ln(f(x))}$ is smooth $p$- times differentiable function representing the refractive index $n$ or the optical path $OP$, and $f^{(m)}(x)$ is the $m^{th}$ derivative of $f(x)$. The $G_p^{(\alpha)}(x)$ polynomials are not orthogonal in general, however, they obey similar relations to the Laguerre polynomials:

$$G_p^{(\alpha)}(x) = G_p^{(\alpha+1)}(x) - pG_{p-1}^{(\alpha+1)}(x) = p!\sum_{j=0}^{\beta}\binom{\beta}{j}\frac{(-1)^j}{(p-j)!}G_{p-j}^{(\alpha+\beta)}(x)$$

$$G_p^{(\alpha+1)}(x) = p!\sum_{j=0}^{p}\frac{1}{j!}G_j^{(\alpha)}(x); \quad G_p^{(\alpha)}(x) = p!\sum_{j=0}^{p}\binom{\alpha+p-\beta}{p-j}\frac{1}{j!}G_j^{(\beta-j)}(x)$$

$$G_{p+1}^{(\alpha)}(x) = (2p+1+\alpha)G_p^{(\alpha)}(x) - p(p+\alpha)G_{p-1}^{(\alpha)}(x) + p!\sum_{m=0}^{p+1}\binom{p+\alpha}{p-(m-1)}\frac{x^m f^{(m)}(x)}{(m-1)!}$$

In terms of these polynomials the $p$ order dispersion (POD), for $p \geq 2$ can be written as:

$$POD = \frac{1}{c}(-1)^p\left(\frac{\lambda}{2\pi c}\right)^{p-1} G_p^{(-2)}(\lambda) = \frac{1}{c}(-1)^p\left(\frac{\lambda}{2\pi c}\right)^{p-1}\left(G_p^{(-1)}(\lambda) - pG_{p-1}^{(-1)}(\lambda)\right)$$

The group delay for $p = 1$ can be written as: $GD = \frac{1}{c}\left(f(\lambda) - G_1^{(-1)}(\lambda)\right)$. In particular, the chromatic dispersion transformation to wavelength space can be seen as a generalized Laguerre transformation or scaled generalized Lah transformation. The forward $\mathcal{L}g^{(\alpha)}$ and inverse $\mathcal{L}g^{(\alpha)^{-1}}$ Laguerre transform are defined as [20, 22]:

$$u_i = \mathcal{L}g^{(2)}(x) = \sum_{j=0}^{i}\binom{i+2}{i-j}\frac{i!}{j!}x_j; \quad x_i = \mathcal{L}g^{(2)^{-1}}(u) = \sum_{j=0}^{i}(-1)^{i-j}\binom{i+2}{i-j}\frac{i!}{j!}u_j$$

Further, the Laguerre and the Lah transforms are connected as: $\mathcal{L}g^{(2)}(x) = P^{(3)}(x)\mathcal{L}(x)$, where $P^{(\beta)} = \binom{i+\beta-j-1}{i-j}\frac{i!}{j!}$ is the permutation matrix. The Laguerre polynomial $L_i^{(-1)}(x)$, for $i \geq 0$ also represents the Lah expansion [23]. While for most of the functions of interest $f(x)$, the Taylor series of the phase expansion will converge, the Taylor series are not guaranteed, in general, to converge. Additionally, in general the polynomials $G_p^{(\alpha)}$ are not orthogonal. However, we can obtain a rough estimate for the radius of convergence from the limit:

$$R = \lim_{p\to\infty}\left|\frac{c_{p+1}}{c_p}\right| \cong \lim_{p\to\infty}\underbrace{\left(1 + \frac{\lambda^{p+1}\frac{\partial^{p+1}}{\partial\lambda^{p+1}}f(\lambda)}{\sum_{m=0}^{p}\mathcal{B}(p,m)\lambda^m\frac{\partial^m}{\partial\lambda^m}f(\lambda)}\right)}_{\epsilon(\omega_0)}\left(\frac{\lambda}{2\pi c}\right)\bigg|_{\omega_0}|\omega - \omega_0| < 1$$

As a thumb rule a convergence could be expected for $|\omega - \omega_0| < \omega_0/\epsilon(\omega_0)$, i.e. $|\lambda| < \frac{\epsilon(\lambda_0)}{1+\epsilon(\lambda_0)}\lambda_0$, $|\Delta\lambda| \sim \lambda_0$.

As an illustration for the importance of considering the higher dispersion orders, as well as, the convenience and speed, of using the polynomial dispersion expressions, we consider first, the chromatic dispersion of $CaF_2$ material. For this purpose, we use the analytic Sellmeier equation for the refractive index $n(\lambda)$ in the spectral range $0.15\mu m - 12\mu m$ [24]. The refractive index derivatives are needed to the order of the considered dispersion and they have to be evaluated once. This can be done efficiently fully numerically. We perform a numerical symbolic evaluation of the derivatives of the refractive index with followed value evaluation at each wavelength. The computation time is 1s. The first 10 chromatic dispersion orders are shown on a bi-log plot [25], Fig. 1 A. The effect on the pulse broadening is significant when $\frac{POD}{\tau_L^p}$ is not negligible, where $\tau_L$ is the pulse length. The effect of the pulse broadening is shown on Fig. 1 B for single- cycle broadband pulses centered at $\lambda_L = 300nm$ for propagation inside $CaF_2$ material of thickness $L = 50\mu m$, $\lambda_L = 1500nm$ for $L = 5mm$, and $\lambda_L = 4000nm$ for $L = 5mm$.

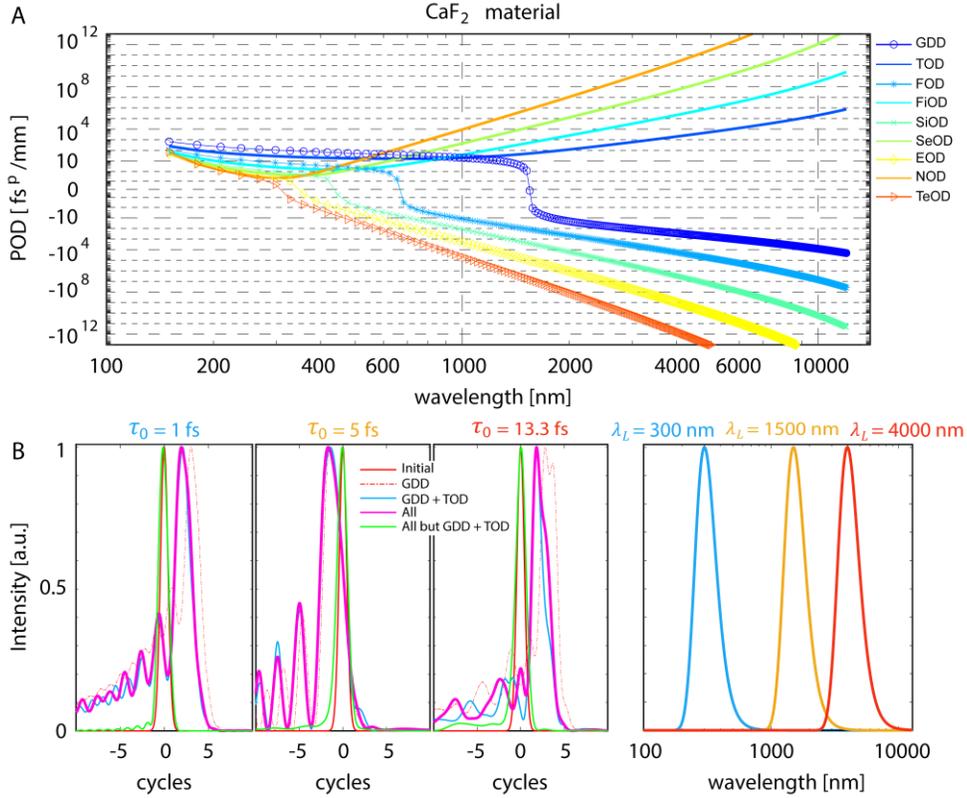

**Figure 1.** A) Chromatic dispersions of order $p$ up to the 10$^{th}$ order for $CaF_2$ material, where $2 \leq p \leq 10$. The long-dashed lines show the numbered tick positions. B) Normalized intensity as function of time (in cycles $\tau_0$), showing the distortions of single-cycle pulses $\tau_0$ at $\lambda_L = 300nm$ after propagation inside $CaF_2$ material of thickness $L = 50\mu m$, $\lambda_L = 1500nm$ for $L = 5mm$ and $\lambda_L = 4000nm$ for $L = 5mm$, due to GDD in dashed red line, GDD and TOD in blue and all orders up the 10$^{th}$ in solid violet. The effect of the higher orders $4 \leq p \leq 10$ only, is shown in green. The right-hand side shows the spectral intensity content of the considered pulses in matching colors.

The pulses have significant spectral content and the wavelength dependent chromatic dispersions are considered when calculating the pulse spreading via Fourier transforms of the waveforms, Fig. 1 B. The high orders dispersions are lower in the UV spectral region and the main pulse broadening is due to the second and third orders, however, the distortions due to the higher dispersion orders are not negligible. The high dispersion orders are significantly more pronounced in the Mid-IR spectral region. The same pattern with slightly larger values is observed for the fused silica-based glasses that are predominantly used in the fiber telecommunication. At the zero GDD dispersion wavelength, long $ps$ narrow-bandwidth pulses are needed to avoid significant spread of the pulses, where the pulse spread is dominated by the TOD, yet higher dispersion orders are also non-negligible.

A general trend for the refractive index away from absorption peaks, is: smooth monotonically decreasing functional form with an inflection point usually in the IR, where is located the zero GDD dispersion wavelength. This behavior gives positive odd order derivatives of the refractive index and step-like behavior with positive and negative values for the even order derivatives.

Next, we consider the chromatic dispersion of a grating compressor with phase $\varphi(\lambda) = \frac{\omega}{c} OP(\lambda)$ and total optical path $OP(\lambda) = 2L(1 - (m\sigma\lambda - sin(\theta))^2)^{\frac{1}{2}}$, where $L$ is the grating separation, $\sigma$ is the grating groove density, $m$ is the diffraction order and $\theta$ is the incident angle [10]. This can be done efficiently fully numerically. First, we perform numerical symbolic evaluation of the derivatives of the optical path with followed value evaluation at each wavelength. The first 10 chromatic dispersion orders are shown on a bi-log plot, Fig. 2 A for $L = 30cm$, $\sigma = 1250\ lines/mm$, $m = 1$ and $\theta = 30deg$. The computation time is 1s.

The higher dispersion orders are large and even a single fourth-order can distort an ultrashort pulse, which makes the grating-compressors not always suitable for compensation of the linear or non-linear phase of temporally and spectrally broadened pulses due to material. Such situation is also encountered in attosecond pulse compression using gratings, where the pulses may not always be fully compensated due to the mismatch with the higher dispersion orders of the grating systems. While GDD can be matched the higher orders may still persist. In a laser system, the compressors are designed with matching stretchers, however, material dispersion adds higher orders residual dispersion that can remains uncompensated. Then a prism- pair compressor can be employed to reduce further the third order while not affecting the GDD. In addition, the higher orders for the prism-compressors are relatively low. A general trend for the grating compressor phase is monotonically increasing values with negative even path derivatives and positive odd path derivatives.

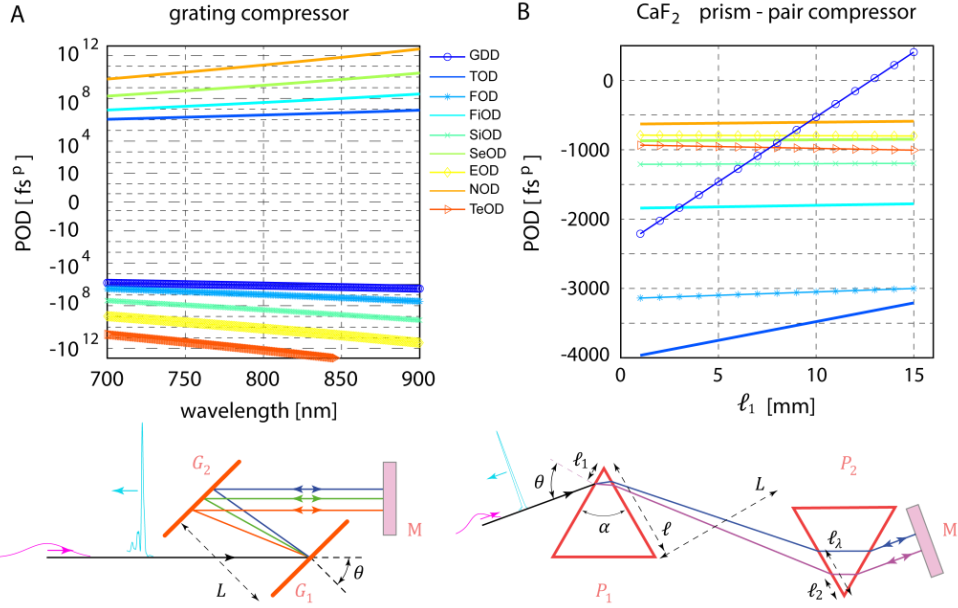

**Figure 2.** Chromatic dispersions of order $p$ up to the 10$^{th}$ order for A) Grating compressor and B) prism - pair compressor estimated at a spectral region of lower values for the high orders of dispersion for the $CaF_2$ material, at $\lambda = 400nm$. The long-dashed lines show the numbered tick positions. Below are shown the schematics of the considered compressors, where $G_i$, $P_i$ and $M$ are the gratings, prisms and the retroreflecting mirror, respectively.

We can also easily evaluate the chromatic dispersion of a prism-pair compressor with phase $\varphi(\lambda) = \frac{\omega}{c} OP(\lambda)$ and total optical path $OP(\lambda)$ including the material dispersion of the prisms [11, 15, 26, 27] for a model considering the finite spectral bandwidth of the pulse [25]: $OP(\lambda) = 2\left(\ell_1 \frac{n^2 sin(\alpha)}{A(\lambda)} + \frac{L}{B(\lambda)} + \ell sin(\theta) + \ell_\lambda \frac{n^2 sin(\alpha) - A(\lambda) sin(\theta)}{D(\lambda)}\right)$, where $\ell_\lambda = \ell_2 + L\left(\frac{C(\lambda_m)}{B(\lambda_m)} - \frac{C(\lambda)}{B(\lambda)}\right) + \ell_1 n \, sin(\alpha) \left(\frac{1}{A(\lambda_m)} - \frac{1}{A(\lambda)}\right)$, $D(\lambda) = \sqrt{(n^2 - sin^2(\theta))}$, $A(\lambda) = cos(\alpha) D(\lambda) + sin(\alpha) sin(\theta)$, $B(\lambda) = \sqrt{1 - C^2(\lambda)}$, $C(\lambda) = sin(\alpha) D(\lambda) - cos(\alpha) sin(\theta)$. The first 10 chromatic dispersion orders are shown on Fig. 2 B for prism of length $\ell = 30mm$, normal distance between the prisms $L = 30cm$, central wavelength $\lambda_0 = 400nm$, bandwidth $\Delta\lambda = 30nm$, apex angle of the prisms $\alpha = 69.9 \, deg$, insertion depth of the first prism $\ell_1 = 1 - 15mm$, insertion depth for the second prism $\ell_2 = 1mm$ for the shortest wavelength $\lambda_m = \lambda_0 - \Delta\lambda/2$, and at Brewster incidence angle $\theta = atan(n(\lambda_0))$. The parameter space is large and one can keep certain order of the chromatic dispersion fixed, while adjusting another. We design the compressor with prisms made out of $CaF_2$ material for the UV spectral region, where the high order chromatic dispersions $p > 3$ are relatively low for the prisms' material. This can be done efficiently fully numerically. First, we perform a numerical symbolic evaluation of the derivatives of the optical path with followed value evaluation as a function of the prism insertion depth $\ell_1$. The computation time is 10s.

Lastly, we estimate the performance of a photonic bandgap- like hollow- core fiber with revolver type substructure [28]. These photonic anti-resonance waveguides have been envisioned for various applications including replacement for the state-of-the-art telecommunication fibers. We evaluate the chromatic dispersion of such fiber with complex wave vector $k(\lambda) = \frac{\omega}{c} n_{eff}^{i,j}(\lambda)$ and effective refractive index $n_{eff}^{1,1}$ for the hybrid mode $HE_{11}$ that is expected to be launched by a linearly polarized gaussian laser mode. The real part of the effective refractive index for the hybrid modes $HE_{ij}$ to first order can be written as: $\Re e(n_{eff}^{i,j}) \cong n_0 - \frac{1}{2n_0}\left(\frac{u_{ij}}{k_0 R}\right)^2 - \frac{u_{ij}^2}{2(k_0 n_0 R)^3} \frac{n^2+n_0^2}{\sqrt{n^2-n_0^2}} cot(\phi)$, where $n_0$ is the refractive index of the hollow core, $n$ is the refractive index of the fiber material, $k_0 = 2\pi/\lambda$, $R$ is the radius of the inner hollow core, $\phi = k_0 \ell \sqrt{n^2 - n_0^2}$, $\ell$ is the thickness of the inner tubes, $u_{11} \cong 2.40483$, and $u_{ij}$ is the $j^{th}$ zero of the Bessel function of first kind $J_{i-1}(x)$ [29]. Hollow- core fibers could potentially offer a larger transferring bit- rate length product considering that light propagates 50% faster in air than in glass that is used in the present-day telecommunication fibers. Further, changing the thickness $\ell$ displaces the zero-dispersion wavelength of the GDD and the even orders, allowing for tuning to or away from the laser wavelength. In addition, these types of fibers can withstand high energies and can maintain a pure polarization state [30].

The first 10 chromatic dispersion orders for the $HE_{11}$ mode are shown on Fig. 3 A for fiber made out of fused silica glass in the spectral range $0.21 \mu m - 3.71 \mu m$ [24], placed in vacuum $n_0 = 1$, with inner radius $R = 30 \mu m$ and wall thickness of $\ell = 1 \mu m$. At these parameters, the major third-order dispersion at the zero-dispersion GDD wavelength is lower compared with the material dispersion of fused silica. However, the higher orders are relatively larger as compared to the material dispersion especially in the UV region. For applications, where the fiber is used as a precursor for starting self-guiding, or for long pulses, these effects are of lesser concern, e.g. in pulse compression, spectral broadening, high harmonic generation, etc.

The general trend for the effective refractive index is: multiple material-like absorptive resonances and monotonically decreasing functional form in-between. This behavior gives multiple zero GDD dispersion wavelengths at the inflection points. The odd orders are strictly positive, while the even, show step-like behavior, change sign and diverge to plus or minus infinity. The GDD is positive on the left side of the zero GDD dispersion point and negative on the right side, thus giving the fortunate prospect for self-compression to few cycle pulse duration for spectrally broadened pulses inside of the gas filled fiber or material. This is especially favorable in the VIS- Mid IR, since the zero- dispersion can be shifted. The major dispersion orders are lower, while the higher orders may have larger values than the material dispersion.

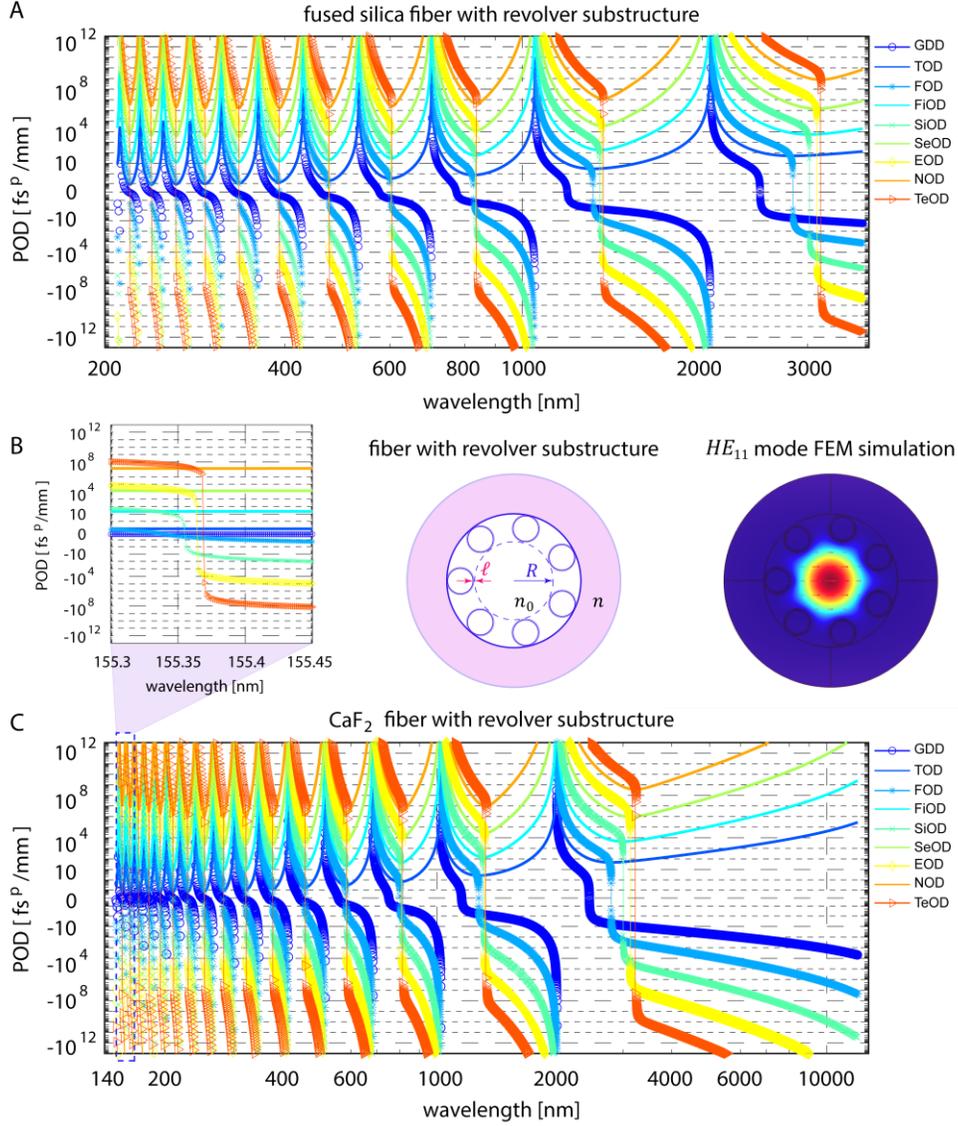

**Figure 3.** Chromatic dispersions of order $p$ up to the $10^{th}$ order for hollow core photonic band gap fiber with inner revolver structure made out of A) fused silica and C) $CaF_2$. The long- dashed lines show the numbered tick positions. B) Zoomed section of the dispersion for the fiber in C), showing the closely spaced zeros of the even dispersion orders. On the left are shown schematics of the structure of the considered fiber, as well as a finite element method (FEM) calculation of the fundamental $HE_{11}$ mode. The red arrows indicate the electric field polarization of the mode.

Additionally, we assess a novel class of non- fused silica - based fibers, made out of $CaF_2$ material in the spectral range $0.15 \mu m - 12 \mu m$. Fig. 3 C shows the first 10 orders of the chromatic dispersions evaluated at the same parameters as discussed above. We observe similar behavior with lower dispersion values. In the deep UV, the zero dispersion wavelengths for the even orders are very close together e.g. within $0.05 nm$ near $155 nm$, Fig. 3 B. At the same time these chromatic dispersion values are divergent away from the zero dispersion wavelengths.

While these classes of fibers might not compete at the moment with the telecommunication fibers in terms of absorption at the telecommunication bands, they offer unique and distinctive advantages. The numerical calculations are more involved at the evaluation of the numerical values of the derivatives. The elapsed time is about 1s per wavelength.

**Conclusion**

We have derived a general analytic expression for the high order chromatic dispersion, due to the $k$ vector or phase $\varphi$ dependence on the wavelength that is valid to infinity. Additionally, we identify polynomials and recursion relations associated with the chromatic dispersion orders, and draw analogy to the generalized Lah and Laguerre transformations. The recursive relationships for the polynomials $G_p^{(\alpha)}$ can be of practical use, when the phase derivatives are repetitive. Further, we have given explicitly the dispersion terms to the 10$^{th}$ order where they can be printed on a paper. These formulas are applicable for material dispersion, compressors, stretchers, waveguides, and any other type of known frequency-dependent phase. As an illustration, we have evaluated the first ten dispersion orders for $CaF_2$ material, for a prism-pair, and reflection grating compressors and photonic anti-resonant fiber. The higher orders of the chromatic dispersion are essential when the propagation lengths are long and/or the spectral content of the pulses is large.

Finally, for phases given as discrete data, the efficiency can be increased by first fitting the data to smooth differentiable functions and proceeding in the usual manner. The computation time decreases significantly with the simplicity of the functions. For more complex systems, parallel computing is very practical. Further, evaluation of the higher order dispersion at or away from the zero GDD dispersion wavelength, is computationally fast, making the minimization of the leading higher orders an efficient procedure for the particular design.

**Acknowledgements**

This work was supported by ERC grant XSTREAM No716950.

**References**

1. Niizeki N. Single Mode Fiber at Zero-dispersion Wavelength. Topical Meeting on Integrated and Guided Wave Optics; 1968; 1968.

2. Handbook of Fiber Optic Data Communication. In: DeCusatis C (ed). *Handbook of Fiber Optic Data Communication (Fourth Edition)*. Academic Press: Oxford, 2013.

3. Agrawal G. *Nonlinear Fiber Optics (Fifth Edition)*. Academic Press: Boston, 2013.

4. Carson MK. *Alexander Graham Bell: Giving Voice to the World*. Sterling, 2007.

5. Hemati S, Emadi M. A method for analyzing higher-order dispersion in optical fiber. CCECE 2003 - Canadian Conference on Electrical and Computer Engineering. Toward a Caring and Humane Technology (Cat. No.03CH37436); 2003 4-7 May 2003; 2003. p. 289-292 vol.281.

6. ITU standard G.655 (11/09), Characteristics of a non-zero dispersion-shifted single-mode optical fibre and cable. *International Telecommunication Union* 2009.

7. Hüttmann E. Distance measuring method (ranging method). *German Patent 768 068* 1940.

8. Cook CE. Pulse Compression-Key to More Efficient Radar Transmission. *Proceedings of the IRE* 1960, **48**(3)**:** 310-316.

9. Strickland D, Mourou G. Compression of amplified chirped optical pulses. *Optics Communications* 1985, **56**(3)**:** 219-221.

10. Treacy EB. Optical Pulse Compression With Diffraction Gratings. *IEEE Journal of Quantum Electronics* 1969, **QE-5**(9)**:** 454-458.

11. Fork RL, Martinez OE, Gordon JP. Negative dispersion using pairs of prisms. *Opt Lett* 1984, **9**(5)**:** 150-152.

12. Heppner J, Kuhl J. Intracavity chirp compensation in a colliding pulse mode-locked laser using thin-film interferometers. *Applied Physics Letters* 1985, **47**(5)**:** 453-455.

13. Yamashita M, Torizuka K, Sato T. A chirp-compensation technique using incident-angle changes of cavity mirrors in a femtosecond pulse laser. *IEEE journal of quantum electronics* 1987, **23**(11)**:** 2005-2007.

14. Szipöcs R, Ferencz K, Spielmann C, Krausz F. Chirped multilayer coatings for broadband dispersion control in femtosecond lasers. *Opt Lett* 1994, **19**(3)**:** 201-203.

15. Jean-Claude Diels WR. *Ultrashort laser pulse phenomena : fundamentals, techniques, and applications on a femtosecond time scale*. Academic Press: Burlington, 2006.

16. Naik PA, Sharma AK. Calculation of Higher Order Group Velocity Dispersion in a Grating Pulse Stretcher / Compressor Using Recursion Method. *Journal of Optics* 2000, **29**(3)**:** 105-113.

17. Popmintchev D, Hernández-García C, Dollar F, Mancuso C, Pérez-Hernández JA, Chen M-C, *et al.* Ultraviolet surprise: Efficient soft x-ray high-harmonic generation in multiply ionized plasmas. *Science* 2015, **350**(6265)**:** 1225-1231.

18. Popmintchev D, Galloway BR, Chen M-C, Dollar F, Mancuso CA, Hankla A, *et al.* Near- and Extended-Edge X-Ray-Absorption Fine-Structure Spectroscopy Using Ultrafast Coherent High-Order Harmonic Supercontinua. *Physical Review Letters* 2018, **120**(9)**:** 093002.

19. Lah I. A new kind of numbers and its application in the actuarial mathematics. *Boletim do Instituto dos Actuários Portugueses* 1954(9)**:** 7–15.

20. Riordan J. *Introduction to Combinatorial Analysis*. Dover Publications, 1958.

21. Kilibarda G, Jovović V. Antichains of Multisets. *Journal of Integer Sequences* 2004, **7**(1).

22. Barry P. Some observations on the Lah and Laguerre transforms of integer sequences. *Journal of Integer Sequences* 2007, **10**(4).

23. Khristo NB. Lah numbers, Laguerre polynomials of order negative one, and the nth derivative of exp(1/x). *Acta Universitatis Sapientiae, Mathematica* 2016, **8**(1)**:** 22-31.

24. *Refractive index database. https://refractiveindex.info/*

25. Webber JBW. A bi-symmetric log transformation for wide-range data. *Measurement Science and Technology* 2012, **24**(2)**:** 027001.

26. Dietel W, Fontaine JJ, Diels JC. Intracavity pulse compression with glass: a new method of generating pulses shorter than 60 fsec. *Opt Lett* 1983, **8**(1)**:** 4-6.

27. Yang Q, Xie X, Kang J, Zhu H, Guo A, Gao Q. Independent and continuous third-order dispersion compensation using a pair of prisms. *High Power Laser Science and Engineering* 2014, **2:** e38.

28. Pryamikov AD, Biriukov AS, Kosolapov AF, Plotnichenko VG, Semjonov SL, Dianov EM. Demonstration of a waveguide regime for a silica hollow - core microstructured optical fiber with a negative curvature of the core boundary in the spectral region > 3.5 μm. *Opt Express* 2011, **19**(2)**:** 1441-1448.

29. Zeisberger M, Schmidt MA. Analytic model for the complex effective index of the leaky modes of tube-type anti-resonant hollow core fibers. *Scientific Reports* 2017, **7**(1)**:** 11761.

30. Taranta A, Numkam Fokoua E, Abokhamis Mousavi S, Hayes JR, Bradley TD, Jasion GT, *et al.* Exceptional polarization purity in antiresonant hollow-core optical fibres. *Nature Photonics* 2020, **14**(8)**:** 504-510.

**Appendix A**

The refractive index $n(\omega)$ and the optical path $OP(\omega)$ are interchangeable in the chromatic dispersion equations. Explicitly the dispersion terms as function of frequency (1) and (2) can be written as:

$$\frac{\partial}{\partial \omega}k(\omega) = \frac{1}{c}\left(n(\omega) + \omega \frac{\partial n(\omega)}{\partial \omega}\right), \frac{d^2}{d\omega^2}k(\omega) = \frac{1}{c}\left(2\frac{\partial n(\omega)}{\partial \omega} + \omega \frac{\partial^2 n(\omega)}{\partial \omega^2}\right)$$

$$\frac{\partial^3}{\partial \omega^3}k(\omega) = \frac{1}{c}\left(3\frac{\partial^2 n(\omega)}{\partial \omega^2} + \omega \frac{\partial^3 n(\omega)}{\partial \omega^3}\right), \frac{\partial^4}{\partial \omega^4}k(\omega) = \frac{1}{c}\left(4\frac{\partial^3 n(\omega)}{\partial \omega^3} + \omega \frac{\partial^4 n(\omega)}{\partial \omega^4}\right)$$

**Appendix B**

The refractive index $n(\omega|\lambda)$, and the optical path $OP(\omega|\lambda)$ are interchangeable in the chromatic dispersion equations. Explicitly the derivatives in the wavelength space (3) and (4) can be transformed as:

$$\frac{\partial n(\omega)}{\partial \omega} = -\left(\frac{\lambda}{2\pi c}\right)G_1^{(-1)}(\lambda) = -\left(\frac{2\pi c}{\omega^2}\right)\frac{\partial n(\omega)}{\partial \lambda} = -\left(\frac{\lambda^2}{2\pi c}\right)\frac{\partial n(\lambda)}{\partial \lambda}$$

$$\frac{\partial^2 n(\omega)}{\partial \omega^2} = \frac{\partial}{\partial \omega}\left(\frac{\partial n(\omega)}{\partial \omega}\right) = \left(\frac{\lambda}{2\pi c}\right)^2 G_2^{(-1)}(\lambda) = \left(\frac{\lambda}{2\pi c}\right)^2\left(2\lambda \frac{\partial n(\lambda)}{\partial \lambda} + \lambda^2 \frac{\partial^2 n(\lambda)}{\partial \lambda^2}\right)$$

$$\frac{\partial^3 n(\omega)}{\partial \omega^3} = -\left(\frac{\lambda}{2\pi c}\right)^3 G_3^{(-1)}(\lambda) = -\left(\frac{\lambda}{2\pi c}\right)^3\left(6\lambda \frac{\partial n(\lambda)}{\partial \lambda} + 6\lambda^2 \frac{\partial^2 n(\lambda)}{\partial \lambda^2} + \lambda^3 \frac{\partial^3 n(\lambda)}{\partial \lambda^3}\right)$$

$$\frac{\partial^4 n(\omega)}{\partial \omega^4} = \left(\frac{\lambda}{2\pi c}\right)^4 G_4^{(-1)}(\lambda) = \left(\frac{\lambda}{2\pi c}\right)^4\left(24\lambda \frac{\partial n(\lambda)}{\partial \lambda} + 36\lambda^2 \frac{\partial^2 n(\lambda)}{\partial \lambda^2} + 12\lambda^3 \frac{\partial^3 n(\lambda)}{\partial \lambda^3} + \lambda^4 \frac{\partial^4 n(\lambda)}{\partial \lambda^4}\right)$$

$$\frac{\partial^5 n(\omega)}{\partial \omega^5} = -\left(\frac{\lambda}{2\pi c}\right)^5 G_5^{(-1)}(\lambda) - \left(\frac{\lambda}{2\pi c}\right)^5\left(120\lambda \frac{\partial n(\lambda)}{\partial \lambda} + 240\lambda^2 \frac{\partial^2 n(\lambda)}{\partial \lambda^2} + 120\lambda^3 \frac{\partial^3 n(\lambda)}{\partial \lambda^3} + 20\lambda^4 \frac{\partial^4 n(\lambda)}{\partial \lambda^4} + \lambda^5 \frac{\partial^5 n(\lambda)}{\partial \lambda^5}\right)$$

$$\frac{\partial^6 n(\omega)}{\partial \omega^6} = \left(\frac{\lambda}{2\pi c}\right)^6 G_6^{(-1)}(\lambda) = \left(\frac{\lambda}{2\pi c}\right)^6\left(720\lambda \frac{\partial n(\lambda)}{\partial \lambda} + 1800\lambda^2 \frac{\partial^2 n(\lambda)}{\partial \lambda^2} + 1200\lambda^3 \frac{\partial^3 n(\lambda)}{\partial \lambda^3} + 300\lambda^4 \frac{\partial^4 n(\lambda)}{\partial \lambda^4} + 30\lambda^5 \frac{\partial^5 n(\lambda)}{\partial \lambda^5} + \lambda^6 \frac{\partial^6 n(\lambda)}{\partial \lambda^6}\right)$$

$$\frac{\partial^7 n(\omega)}{\partial \omega^7} = -\left(\frac{\lambda}{2\pi c}\right)^7 G_7^{(-1)}(\lambda) = -\left(\frac{\lambda}{2\pi c}\right)^7\left(5040\lambda \frac{\partial n(\lambda)}{\partial \lambda} + 15120\lambda^2 \frac{\partial^2 n(\lambda)}{\partial \lambda^2} + 12600\lambda^3 \frac{\partial^3 n(\lambda)}{\partial \lambda^3} + 4200\lambda^4 \frac{\partial^4 n(\lambda)}{\partial \lambda^4} + 630\lambda^5 \frac{\partial^5 n(\lambda)}{\partial \lambda^5} + 42\lambda^6 \frac{\partial^6 n(\lambda)}{\partial \lambda^6} + \lambda^7 \frac{\partial^7 n(\lambda)}{\partial \lambda^7}\right)$$

$$\frac{\partial^8 n(\omega)}{\partial \omega^8} = \left(\frac{\lambda}{2\pi c}\right)^8 G_8^{(-1)}(\lambda) = \left(\frac{\lambda}{2\pi c}\right)^8\left(40320\lambda \frac{\partial n(\lambda)}{\partial \lambda} + 141120\lambda^2 \frac{\partial^2 n(\lambda)}{\partial \lambda^2} + 141120\lambda^3 \frac{\partial^3 n(\lambda)}{\partial \lambda^3} + 58800\lambda^4 \frac{\partial^4 n(\lambda)}{\partial \lambda^4} + 11760\lambda^5 \frac{\partial^5 n(\lambda)}{\partial \lambda^5} + 1176\lambda^6 \frac{\partial^6 n(\lambda)}{\partial \lambda^6} + 56\lambda^7 \frac{\partial^7 n(\lambda)}{\partial \lambda^7} + \lambda^8 \frac{d^8 n(\omega)}{d\lambda^8}\right)$$

$$\frac{\partial^9 n(\omega)}{\partial \omega^9} = -\left(\frac{\lambda}{2\pi c}\right)^9 G_9^{(-1)}(\lambda) = -\left(\frac{\lambda}{2\pi c}\right)^9 \left(362880\lambda \frac{\partial n(\lambda)}{\partial \lambda} + 1451520\lambda^2 \frac{\partial^2 n(\lambda)}{\partial \lambda^2} + 1693440\lambda^3 \frac{\partial^3 n(\lambda)}{\partial \lambda^3} + 846720\lambda^4 \frac{\partial^4 n(\lambda)}{\partial \lambda^4} + 211680\lambda^5 \frac{\partial^5 n(\lambda)}{\partial \lambda^5} + 28224\lambda^6 \frac{\partial^6 n(\lambda)}{\partial \lambda^6} + 2016\lambda^7 \frac{\partial^7 n(\lambda)}{\partial \lambda^7} + 72\lambda^8 \frac{\partial^8 n(\lambda)}{\partial \lambda^8} + \lambda^9 \frac{d^9 n(\lambda)}{d\lambda^9}\right)$$

$$\frac{\partial^{10} n(\omega)}{\partial \omega^{10}} = \left(\frac{\lambda}{2\pi c}\right)^{10} G_{10}^{(-1)}(\lambda) = \left(\frac{\lambda}{2\pi c}\right)^{10} \left(3628800\lambda \frac{\partial n(\lambda)}{\partial \lambda} + 16329600\lambda^2 \frac{\partial^2 n(\lambda)}{\partial \lambda^2} + 21772800\lambda^3 \frac{\partial^3 n(\lambda)}{\partial \lambda^3} + 12700800\lambda^4 \frac{\partial^4 n(\lambda)}{\partial \lambda^4} + 3810240\lambda^5 \frac{\partial^5 n(\lambda)}{\partial \lambda^5} + 635040\lambda^6 \frac{\partial^6 n(\lambda)}{\partial \lambda^6} + 60480\lambda^7 \frac{\partial^7 n(\lambda)}{\partial \lambda^7} + 3240\lambda^8 \frac{\partial^8 n(\lambda)}{\partial \lambda^8} + 90\lambda^9 \frac{\partial^9 n(\lambda)}{\partial \lambda^9} + \lambda^{10} \frac{\partial^{10} n(\lambda)}{\partial \lambda^{10}}\right)$$

**Appendix C**

The refractive index $n(\omega|\lambda)$, and the optical path $OP(\omega|\lambda)$ are interchangeable in the chromatic dispersion equations. Explicitly the dispersion terms in the frequency and wavelength space (5) and (6) can be written as:

$$GD = \frac{\partial}{\partial \omega} k(\omega) = \frac{1}{c}\left(n(\omega) + \omega \frac{dn(\omega)}{d\omega}\right) = \frac{1}{c}\left(n(\lambda) - \lambda \frac{dn(\lambda)}{d\lambda}\right) = v_{gr}^{-1}$$

The group refractive index is defined as: $n_g = c v_{gr}^{-1}$

$$GDD = \frac{\partial^2}{\partial \omega^2} k(\omega) = \frac{1}{c}\left(2\frac{\partial n(\omega)}{\partial \omega} + \omega \frac{\partial^2 n(\omega)}{\partial \omega^2}\right) = \frac{1}{c}\left(\frac{\lambda}{2\pi c}\right) G_2^{(-2)}(\lambda) = \frac{1}{c}\left(\frac{\lambda}{2\pi c}\right)\left(\lambda^2 \frac{\partial^2 n(\lambda)}{\partial \lambda^2}\right)$$

$$TOD = \frac{\partial^3}{\partial \omega^3} k(\omega) = \frac{1}{c}\left(3\frac{\partial^2 n(\omega)}{\partial \omega^2} + \omega \frac{\partial^3 n(\omega)}{\partial \omega^3}\right) = -\frac{1}{c}\left(\frac{\lambda}{2\pi c}\right)^2 G_3^{(-2)}(\lambda) = -\frac{1}{c}\left(\frac{\lambda}{2\pi c}\right)^2 \left(3\lambda^2 \frac{\partial^2 n(\lambda)}{\partial \lambda^2} + \lambda^3 \frac{\partial^3 n(\lambda)}{\partial \lambda^3}\right)$$

$$FOD = \frac{\partial^4}{\partial \omega^4} k(\omega) = \frac{1}{c}\left(4\frac{\partial^3 n(\omega)}{\partial \omega^3} + \omega \frac{\partial^4 n(\omega)}{\partial \omega^4}\right) = \frac{1}{c}\left(\frac{\lambda}{2\pi c}\right)^3 G_4^{(-2)}(\lambda) = \frac{1}{c}\left(\frac{\lambda}{2\pi c}\right)^3 \left(12\lambda^2 \frac{\partial^2 n(\lambda)}{\partial \lambda^2} + 8\lambda^3 \frac{\partial^3 n(\lambda)}{\partial \lambda^3} + \lambda^4 \frac{\partial^4 n(\lambda)}{\partial \lambda^4}\right)$$

$$FiOD = \frac{\partial^5}{\partial \omega^5} k(\omega) = \frac{1}{c}\left(5\frac{\partial^4 n(\omega)}{\partial \omega^4} + \omega \frac{\partial^5 n(\omega)}{\partial \omega^5}\right) = -\frac{1}{c}\left(\frac{\lambda}{2\pi c}\right)^4 G_5^{(-2)}(\lambda) = -\frac{1}{c}\left(\frac{\lambda}{2\pi c}\right)^4 \left(60\lambda^2 \frac{\partial^2 n(\lambda)}{\partial \lambda^2} + 60\lambda^3 \frac{\partial^3 n(\lambda)}{\partial \lambda^3} + 15\lambda^4 \frac{\partial^4 n(\lambda)}{\partial \lambda^4} + \lambda^5 \frac{\partial^5 n(\lambda)}{\partial \lambda^5}\right)$$

$$SiOD = \frac{\partial^6}{\partial \omega^6} k(\omega) = \frac{1}{c}\left(6\frac{\partial^5 n(\omega)}{\partial \omega^5} + \omega \frac{\partial^6 n(\omega)}{\partial \omega^6}\right) = \frac{1}{c}\left(\frac{\lambda}{2\pi c}\right)^5 G_6^{(-2)}(\lambda) = \frac{1}{c}\left(\frac{\lambda}{2\pi c}\right)^5 \left(360\lambda^2 \frac{\partial^2 n(\lambda)}{\partial \lambda^2} + 480\lambda^3 \frac{\partial^3 n(\lambda)}{\partial \lambda^3} + 180\lambda^4 \frac{\partial^4 n(\lambda)}{\partial \lambda^4} + 24\lambda^5 \frac{\partial^5 n(\lambda)}{\partial \lambda^5} + \lambda^6 \frac{\partial^6 n(\lambda)}{\partial \lambda^6}\right)$$

$$SeOD = \frac{\partial^7}{\partial \omega^7} k(\omega) = \frac{1}{c}\left(7\frac{\partial^6 n(\omega)}{\partial \omega^6} + \omega \frac{\partial^7 n(\omega)}{\partial \omega^7}\right) = -\frac{1}{c}\left(\frac{\lambda}{2\pi c}\right)^6 G_7^{(-2)}(\lambda) = -\frac{1}{c}\left(\frac{\lambda}{2\pi c}\right)^6 \left(2520\lambda^2 \frac{\partial^2 n(\lambda)}{\partial \lambda^2} + 4200\lambda^3 \frac{\partial^3 n(\lambda)}{\partial \lambda^3} + 2100\lambda^4 \frac{\partial^4 n(\lambda)}{\partial \lambda^4} + 420\lambda^5 \frac{\partial^5 n(\lambda)}{\partial \lambda^5} + 35\lambda^6 \frac{\partial^6 n(\lambda)}{\partial \lambda^6} + \lambda^7 \frac{\partial^7 n(\lambda)}{\partial \lambda^7}\right)$$

$$EOD = \frac{\partial^8}{\partial \omega^8} k(\omega) = \frac{1}{c}\left(8\frac{\partial^7 n(\omega)}{\partial \omega^7} + \omega\frac{\partial^8 n(\omega)}{\partial \omega^8}\right) = \frac{1}{c}\left(\frac{\lambda}{2\pi c}\right)^7 G_8^{(-2)}(\lambda) =$$

$$\frac{1}{c}\left(\frac{\lambda}{2\pi c}\right)^7 \left(20160\lambda^2 \frac{\partial^2 n(\lambda)}{\partial \lambda^2} + 40320\lambda^3 \frac{\partial^3 n(\lambda)}{\partial \lambda^3} + 25200\lambda^4 \frac{\partial^4 n(\lambda)}{\partial \lambda^4} + 6720\lambda^5 \frac{\partial^5 n(\lambda)}{\partial \lambda^5} + 840\lambda^6 \frac{\partial^6 n(\lambda)}{\partial \lambda^6} + 48\lambda^7 \frac{\partial^7 n(\lambda)}{\partial \lambda^7} + \lambda^8 \frac{\partial^8 n(\lambda)}{\partial \lambda^8}\right)$$

$$NOD = \frac{\partial^9}{\partial \omega^9} k(\omega) = \frac{1}{c}\left(9\frac{\partial^8 n(\omega)}{\partial \omega^8} + \omega\frac{\partial^9 n(\omega)}{\partial \omega^9}\right) = -\frac{1}{c}\left(\frac{\lambda}{2\pi c}\right)^8 G_9^{(-2)}(\lambda) =$$

$$-\frac{1}{c}\left(\frac{\lambda}{2\pi c}\right)^8 \left(181440\lambda^2 \frac{\partial^2 n(\lambda)}{\partial \lambda^2} + 423360\lambda^3 \frac{\partial^3 n(\lambda)}{\partial \lambda^3} + 317520\lambda^4 \frac{\partial^4 n(\lambda)}{\partial \lambda^4} + 105840\lambda^5 \frac{\partial^5 n(\lambda)}{\partial \lambda^5} + 17640\lambda^6 \frac{\partial^6 n(\lambda)}{\partial \lambda^6} + 1512\lambda^7 \frac{\partial^7 n(\lambda)}{\partial \lambda^7} + 63\lambda^8 \frac{\partial^8 n(\lambda)}{\partial \lambda^8} + \lambda^9 \frac{\partial^9 n(\lambda)}{\partial \lambda^9}\right)$$

$$TeOD = \frac{\partial^{10}}{\partial \omega^{10}} k(\omega) = \frac{1}{c}\left(10\frac{\partial^9 n(\omega)}{\partial \omega^9} + \omega\frac{\partial^{10} n(\omega)}{\partial \omega^{10}}\right) = \frac{1}{c}\left(\frac{\lambda}{2\pi c}\right)^9 G_{10}^{(-2)}(\lambda) =$$

$$\frac{1}{c}\left(\frac{\lambda}{2\pi c}\right)^9 \left(1814400\lambda^2 \frac{\partial^2 n(\lambda)}{\partial \lambda^2} + 4838400\lambda^3 \frac{\partial^3 n(\lambda)}{\partial \lambda^3} + 4233600\lambda^4 \frac{\partial^4 n(\lambda)}{\partial \lambda^4} + 1693440\lambda^5 \frac{\partial^5 n(\lambda)}{\partial \lambda^5} + 352800\lambda^6 \frac{\partial^6 n(\lambda)}{\partial \lambda^6} + 40320\lambda^7 \frac{\partial^7 n(\lambda)}{\partial \lambda^7} + 2520\lambda^8 \frac{\partial^8 n(\lambda)}{\partial \lambda^8} + 80\lambda^9 \frac{\partial^9 n(\lambda)}{\partial \lambda^9} + \lambda^{10} \frac{\partial^{10} n(\lambda)}{\partial \lambda^{10}}\right)$$

Where $c = 299\,792\,458$ m/s is the speed of light in vacuum.